# Enabling Micro-level Demand-Side Grid Flexiblity in Resource Constrained Environments

Ponce de Leon Barido, D.[1], Rosa, J.[2], Suffian, S.[3], Brewer, E.[2], Kammen, D.M[1,4]

Corresponding Author: diego.leon@berkeley.edu

[1] The Energy and Resources Group, University of California Berkeley
[2] Electrical Engineering and Computer Sciences, University of California Berkeley
[3] Department of Electrical and Computer Engineering, University of Villanova
[4] Goldman School of Public Policy, University of California Berkeley

*Abstract*—The increased penetration of uncertain and variable renewable energy presents various resource and operational electric grid challenges. Micro-level (household and small commercial) demand-side grid flexibility could be a cost-effective strategy to integrate high penetrations of wind and solar energy, but literature and field deployments exploring the necessary information and communication technologies (ICTs) are scant. This paper presents an exploratory framework for enabling information driven grid flexibility through the Internet of Things (IoT), and a proof-of-concept wireless sensor gateway (FlexBox) to collect the necessary parameters for adequately monitoring and actuating the micro-level demand-side. In the summer of 2015, thirty sensor gateways were deployed in the city of Managua (Nicaragua) to develop a baseline for a near future small-scale demand response pilot implementation. FlexBox field data has begun shedding light on relationships between ambient temperature and load energy consumption, load and building envelope energy efficiency challenges, latency communication network challenges, and opportunities to engage existing demand-side user behavioral patterns. Information driven grid flexibility strategies present great opportunity to develop new technologies, system architectures, and implementation approaches that can easily scale across regions, incomes, and levels of development.

## I. INTRODUCTION

The penetration of uncertain and variable renewable energy is now occurring across many regions, incomes, and levels of development. In the immediate future, countries such as Uruguay are expected to produce 35% of their generation from wind energy alone (2016), Kenya expects 300 MW of wind to come online in 2016, Thailand will develop 3 gigawatts (GW) of rooftop and village based solar projects (2021), and Africa's Clean Energy Corridor should significantly increase the penetration of renewable energy in the continent [1][2]. In Central America, Costa Rica has produced up to 100% of its generation from renewable resources (~25% without large hydropower), and Nicaragua produces ~40% of it's total generation from non-large hydropower renewable resources. Indeed, some research suggests that between 2015 and 2040 approximately $US12.2 trillion will be invested in global power generation, with two thirds of the total being dedicated to renewable energy, and with the great majority (78%) of this investment occurring in emerging economies [3].

Similarly to future trends in power generation, future growth in urbanization will also mainly occur in cities of the rising south. UN Habitat reports that in the past decade, the urban population in emerging economies grew on average 1.2 million people *per week*, with Asia adding 0.8 million new urban dwellers every week, followed by Africa (0.23 million/week), and Latin American and the Caribbean (0.15 million/week) [4]. It is expected that seven out of ten people will be living in cities by 2050 [4][5]. Similarly, the growth in the use of cellphones and smartphones is also unprecedented. Currently there are more active mobile connections (7.5 billion and 3.7 billion unique connections) than people in the world (7 billion), with penetration rates being large even in low-income economies (89 subscriptions per 100 people) [6][7]. Although currently low, the number of 3G/4G users is expected to double by 2020 (2.5 billion users) [8]. With most renewable energy, urbanization and connectivity growth in the coming decades occurring in low and lower-middle income countries, it is becoming increasingly important to understand how to harvest information from resource constrained environments (RCEs) to provide value both to users and urban services (for example, energy, water, transportation, and banking) [5][9][10].

In this paper we introduce the FlexBox, a wireless sensor gateway and associated suite of sensors that monitors and controls thermostatically controlled loads (TCLs), integrates TCL state information with household-level electricity metering, and combines this information with grid level data in the cloud. We demonstrate the potential for this gateway to provide demand response ancillary services for renewable energy integration in Managua, Nicaragua. Although Nicaragua has the second lowest GDP per capita in the Western Hemisphere, it is also one of the countries in the world with the highest penetration of non-hydropower renewable energy (on an hourly basis, it can produce up to 50% of its generation from wind power alone) [1][11]. This paper first describes related work (Section II), Section III introduces the FlexBox as a part of a system, and Section IV describes a proof of concept demonstration in a laboratory setting. Section V discusses preliminary results and challenges from the technology's first deployment in Nicaragua. We present future opportunities and challenges for networking thermostatically controlled loads in resource-constrained environments in Section VI.

## II. RELATED WORK

Attention towards the actuation of TCLs has grown as the penetration of intermittent renewable energy increases with innovations being made in theoretical frameworks, controlled environment pilot tests, development of new technologies, and field deployments [12][13][14]. In this section, we review theoretical approaches to demand side grid flexibility, also known as demand response, review existing solutions to the actuation of thermostatically controlled loads to provide power grid services, and briefly discuss field deployments available in the literature.

### A. Theoretical Approaches to Demand Response

DR is related to the 'end-use', with electric loads (and users) reducing or shifting their usage in a given time period in response to a price signal, financial incentive, environmental condition or reliability signal [15]. Electric loads exist under three broad categories: *(1)* inflexible/on-demand (lights, television sets, radios, desktop computers), *(2)* deferrable (washers, dryers, dishwashers), *(3)* and flexible (HVAC, EVs, refrigerators, water heaters) [13]. Smart flexible energy loads (TCLs), contain enough local energy to run for an extended period of time, and an intelligent controller can engage with them via direct load control (DLC) to manage energy reserves without significantly inhibiting operation [16]. Furthermore, TCLs can be controlled for the purpose of curtailment, substitution, storage, and/or load shifting [17]. The theoretical approach to demand response has motivated a wide body of work that seeks to show that large aggregations of TCLs can be used both to bid into grid related ancillary service markets for profit as well as to maintain reliable power system operations [18][19][20][21][22]. Detailed end-use models explore associated uncertainties in aggregating TCLs, algorithmic bidding approaches toggle load switch controllers for managing wind forecast error and reducing external balancing penalties, and control and differential equation approaches for modeling the effect of broadcasting signals for TCL set point adjustments [17][23][24][25]. In general, room temperature, inside temperature (of a room, or inside a refrigerator, for example), power consumption, and TCL characteristics (resistance, capacitance, and wall thickness, for example) are all used for the design of a smart controller [20]. Although complex and highly detailed, many of these models make simplifying assumptions that could significantly affect model choice and development including the possibility of heterogeneity across micro-climates in urban areas, insulation of houses (or buildings) where TCLs reside, load efficiency, the size (and varying thermal mass) of TCLs, and random user behavioral patterns that can drive TCL cycling. Furthermore, data acquisition devices (sensors and power meters, for example) and communications platforms and protocols are discussed abstractly, with most of the effort being directed towards the development of theoretical frameworks to further enable greater grid flexibility.

### B. Available Sensing, Actuation and Control Solutions

There exists a panoply of solutions for networking sensors and plug load devices in higher-income countries, but the literature is scant regarding how sensing infrastructure will grow and be networked in countries with low internet access, and spotty communication networks. In countries with high internet penetration, the monitoring of loads can be done through Wi-Fi enabled large smart home appliances, plug-level monitors (individually installed and programmed, often with proprietary communication protocols), and non-intrusive load monitoring (NILM) [26]. NILM is an alternative to ubiquitous plug level monitoring, only requiring one energy meter to monitor whole house energy consumption, and signal processing for load disaggregation, but to date, this remains mostly a research effort [26]-[35].

Communication protocols for energy reporting and control of devices are primarily designed for local area networks (LAN) (e.g. stacks such as Zigbee or Z-Wave), and APIs, such as OpenADR and GreenButton which are intended for use over the internet [26]. Zigbee's Smart Energy Profile enables low-power device monitoring using 802.15.4 radios and links that support IPv6 through an HTTP interface, and OpenADR 2.0b and GreenButton Connect are XML standards for energy data exchanges between utilities, consumers, and third-party service providers [26]. Differently from Zigbee and IETF 6LowWPAN (which use IEEE 802.15.4) Z-Wave uses a proprietary low-latency transmission communications protocol that uses small data packets at 100kbit/s, operating through a source-routed mesh network that helps the device avoid obstacles and radio dead spots in a multipath environment. Zigbee and Z-Wave differs from OpenADR and GreenButton in that the latter were designed for high-bandwidth network connections and large files sizes, making them less useful for low-power local area networks. Z-wave uses a source-routed mesh network architecture. In addition, Bluetooth (communicating over IEEE 802.15.1) can be used for short-range applications to replace cables for computer peripherals such as mice, keyboards, and printers, but to date has few applications for monitoring and control of electric loads [35][36][37].

In California, Radio Broadcast Data Systems (RBDS) have been recommended as the statewide DR broadcasting signaling standard, and it has been shown that RBDS can be used to broadcast one-way demand response messages with near 100 percent probability using merely just one FM station [38][39][40]. RBDS use a 57 kilohertz subcarrier to transmit over 1 kbts, with data being transmitted in groups of four blocks (26 bits each) [38][39]. All available FM channels (frequencies) within proximity can be used to broadcast signals, with the probability of message reception being dependent on signal strength and the number of message repeats [38]. Since DR applications only add one to two percent of total average transmission station capacity to a channel, FM station contract costs for RBDS are relatively low (hundreds of dollars per month) [38]. A downside, however, is the one-way nature of the FM broadcast.

*C. Research Pilots and Field Deployments*

While there have been a variety of approaches that have been shown to be effective in simulation there are two principal questions that have largely remained unaddressed: *1)* how well do algorithms and loads behave in practice? and *2)* what is the actual size of the resource that is available for demand response in a region or country? Research pilots have investigated the potential of *deferrable* and *flexible* loads to provide grid-flexibility using loads as virtual power plants and exploring opportunities for users to experience energy savings through real time pricing. These pilots have instrumented as few as one and as many as five refrigerators to study real-time behavior of loads under DR [13][40][41]. A few have also developed proprietary thermal-storage eutectic phase-change storage systems that can be controlled [13][40]. Some of the business scenarios explored in these pilots include: (1) the aggregation and market-auctioning of thermal storage 'virtual power plants' (controllable tool kits are *given* to businesses and households for a load aggregator to make a profit through auctioning), (2) 'smart refrigerators' independently taking advantage of real time pricing opportunities (users *buy* a controllable tool kit to take advantage of real time pricing), and (3) incentives for supply following loads.

In California, some research pilots have suggested that a 'thermal storage refrigerator' controlled through a load aggregation framework could have great value and a relatively fast five-year payback period, while others have found that Californian households would only benefit from buying 'controllable tool kits' if real time prices were slightly higher than what they currently are [13][40]. [13] finds that household savings in California would be negligible due to the amount of energy required to freeze and control an actionable phase change material, and [13][40] both find that the amount of savings experienced by a household depends substantially on the pricing tariff. In Denmark, research pilots suggest that the 'micro-payments' provided to users for participating in a load aggregation would be too low (1 to 5 euros/month) and energy cost savings would be too little (1 to seven euros/year) from buying a 'controllable toolkit'. The absence of business potential in Denmark depended heavily on rate structure and other fees that make up a large part of the electricity prices (fixed costs being a large proportion of the electricity bill, rather than variable costs), in addition to refrigerators being much smaller and efficient than Californian refrigerators, and thus, requiring a larger population to take full advantage of virtual storage plants.

Manual DR (manually changing set points, with a switch or controllers, for example), semi-automated DR (automating HVAC or other processes through the use of energy management control systems, with the remainder of a facility under 'human control'), and fully automated DR (automation of an entire facility) are the three most popular ways to implement DR research pilots and field deployments. High data granularity through metering or advanced metering infrastructure (AMI) is essential for all DR implementations to ensure project performance and end-user compliance, financial settlement, and consumer satisfaction (by providing access to data), among other things [42][43][44]. Currently, most AMI deployments interphase with smart metering infrastructure through analog pulse or digital series outputs, as well as metering specific loads. These data are for the most part sent back to an aggregator through existing communications infrastructure such as broadband or wifi [43][44]. With response times that can range from tens of minutes to milliseconds, DLC is an integral part of AMI kits used in pilots and research projects to ensure compliance, as it would be nearly impossible for users to act within some of the shortest time frames [37]. Two-way communications have also been crucial as it allows toggling relays, sending scripts to BEMS, or attachment to a wide assortment of loads or industrial equipment [37]. Network Operation Centers (NOCs), or centralized control servers, host and organize DR and are widely used in commercialized implementations for initiating automatic dispatch notifications, remote control and monitoring of customer loads and generation, and coordination technicians in the field [37].

In the United States, OpenADR is now almost always used as the communications data model of choice to *'facilitate sending and receiving DR signals from a utility or independent system operator to electric customers'* [37][38]. While the OpenADR specification certainly facilitates data exchanges across a variety of stakeholders including consumers, utilities, regional transmission organizations (RTOs), and independent system operators (ISOs), it was primarily designed for high-bandwidth networks rather than low-power local area networks, making it perhaps less useful for smaller research tailored implementations or niche markets. Once the system is in place, providing capacity payments, enabling meter access, facilitating accurate and transparent measurement verification procedures (establishing a baseline, for example), and encouraging aggregation are seen as industry best standards [37].

Research pilots and field deployments are an important next step in realizing the implementation of the smart grid, and future research projects will have to further investigate important aspects of DR implementations including two-way communications costs and/or challenges, and the incorporation of behavior in DR (opening and closing of doors, for example), which plays an important role in TCL cycling. Another important challenge to consider is that AMI and smart metering were not designed with DR or other ancillary services in mind. In California (PG&E), smart meter infrastructure may receive or send several signals per day, with the transmission frequency depending on its position across a mesh network, and hence, does not provide all the functionality that DR aggregators would like when ensuring high standards for project implementation. Furthermore, DLC programs have historically faced end-user challenges including customers becoming frustrated with service interruptions, and often times leaving programs if they are called on too frequently, or not offered sufficient incentives to maintain long-term project participation [44]. Technology innovation in networking and DR technologies needs to consider many of these challenges. At the micro-level, networking technologies and the rapid decline in microprocessors and sensing technology provide ample opportunity to help implement successful socio-technological DR interventions. While a number of initiatives have already begun exploring the challenges of monitoring and controlling loads for DR at micro-level, the development of a scalable control methodology is still a major challenge [38].

## III. System Concept

In January 2015 we used the Open Data Kit platform to survey 230 micro-enterprises with large cooling loads in Managua. A pilot survey was tested with a small group of 20 micro-enterprises, adjustments were made, and a full implementation was performed immediately afterwards. Our surveys and conversations with micro-enterprises (MEs) with large-cooling loads (for example: butcheries, chicken shops, mom & shops, milk and cheese hops) attempted to assess whether a micro-level demand response implementation could be feasible in Nicaragua and touched upon different aspects of a micro-enterprise's business model: income and cost structures, energy related expenditures, daily, monthly and seasonal variations in consumption, perceptions on electric service reliability, perceptions on the quality of service provided by the utility, relationship with loads and appliances, and perceptions on income and micro-enterprise expenditures [45]

The three most salient results from this survey included learning about *(1) Voluntary Load Disruption:* 161 respondents (71% of sample), were already implementing a refrigerator 'energy savings strategy' by turning their refrigerator on or off at different times of the day, *(2) Perceptions on Electricity Service Reliability:* Despite 70% of the MEs experiencing frequent power outages, most were '*satisfied*' (72%) with service reliability (our data, however, registered very low voltages across the geographic spectrum, affecting the performance of certain appliances such as refrigerators), suggesting a high level of acceptance towards loads (and service) being turned off at random, and *(3) High Energy Costs and Perception of Electricity Related Expenditures:* The MEs' main cost concerns were related to high energy prices (US$ 0.33/kWh), with 60% finding their bills 'difficult to pay' (on a scale from 1-4: 'easy' to 'very difficulty to pay') [5]. The objective of the system is to turn everyday TCLs (refrigerators, in this instance) into grid-tied 'batteries' that have the ability to store energy via latent heat, while still being able to perform their intended tasks. The system gathers open access high-resolution grid and weather data, as well as information from micro-level users such as micro-enterprises and homes via surveys and a wireless sensor gateway. Actionable signals and personalized and useful snippets of 'energy efficient' information are developed in the cloud and can be pushed back to users, but understanding the state of an aggregated 'virtual storage plant' (as described above) for DR simulation and control is the primary task of our design and implementation.

With knowledge of previously implemented micro-level DR implementations, and taking into account characteristics and challenges particular to Nicaragua, a system was conceived that could scale across regions and levels of infrastructural development (grid flexibility enabled box: FlexBox). The FlexBox requires intelligence far beyond a power meter; its design must allow for the possibility of using information about household energy consumption, refrigerator energy consumption, refrigerator temperature, refrigerator usage, and room temperature to independently make decisions about turning the refrigerator on and off. Similarly, its design must also allow for the possibility of two-way communications with a load aggregator. These functionalities were not implemented in a vacuum and followed a set of design principles that fit the deployment and project context. The principles surrounding FlexBox design were guided by the needs of all the "users" including: *1) adaptability:* the team of researchers (at the University of California and the Nicaraguan National Engineering University) who will need to develop DR control laws, sensor configurations and management, and data collection and transmission functionalities, *2) modularity:* simple maintenance being performed by a local enumerator without formal training in electronics meant that the system components could be put together and apart with ease, and *3)*

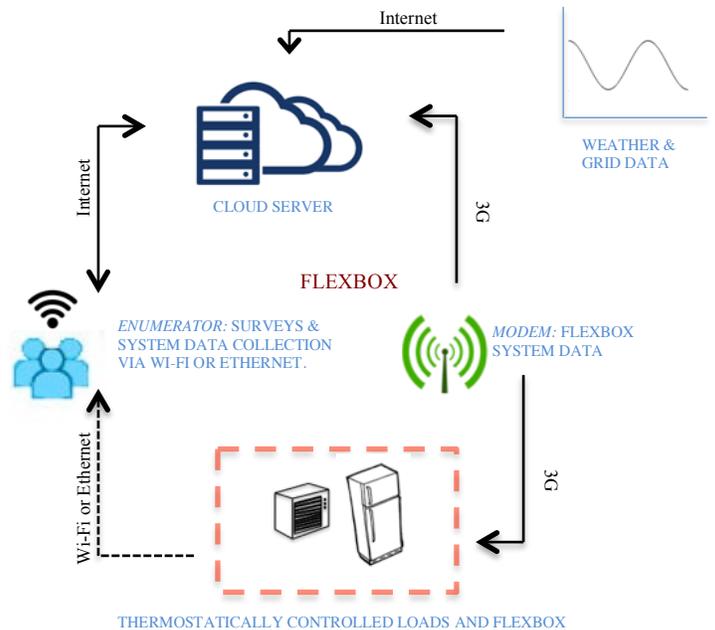

**Figure 1. FlexBox System Concept:** The enumerator downloads new FlexBox software and new surveys from the cloud server. The enumerator also collects data from the FlexBox via Ethernet or Wi-Fi and sends it to the cloud server. A Huawei E3531 modem opens two-way communication streams between the FlexBox and the cloud server, uploading data and downloading updated control laws. Open access grid and weather data are stored in the cloud server as well as an archive of transmitted data.

*user needs and acceptance:* the home or business owners must accept the technology, and at a later stage, receive information about household and load consumption. The second set of design principles came from our motivational research questions. A compromise had to be found between the availability of low-cost sensors and the variables of interest, including human behavior. This set of principles, while dominating the motivation of the project, is often the easiest to satisfy as it is sufficiently under the control of the researchers.

## IV. Proof of Concept Demonstration

The FlexBox is designed for ubiquitous TCL and household sensing, monitoring and load control. In this section we discuss the principles of operation, the hardware and software implementation, and the laboratory tests that were carried out before the field deployment.

### A. Principles of Operation

Our first research pilot in Managua consists of thirty FlexBoxes attached to twenty freezers (micro-enterprises) and ten refrigerators (households) and a centralized server that stores data, performs analyses, and provides control signals. Each FlexBox collects fridge inside temperature, humidity, TCL energy consumption, and total household energy consumption and stores it in a local database. Data is sent over 3G to a centralized server where it is merged with time stamped open access grid and weather data. Statistical and control scripts in the server can run simulations, and when necessary, actionable DR signals can be sent to participating TCLs to either be turned off or return to their normal cycling schedules. This central server also provides web-based tools to export data for off-line analysis, user energy reports, and intuitive visualizations that allow interested parties to easily understand the state of the overall system.

### B. Hardware Implementation

FlexBox processing and data storage is managed by a Raspberry PI B+. This platform provides a full operating system as the development environment which provides a richer feature set familiar to all researchers (UC Berkeley and UNI), which would not be the case if a simpler system were used, such as an mBed or Arduino microcontroller, which use a subset of C++. Four USB ports are used to add and test wireless communications peripherals for local device communications using the Z-Wave protocol, Wi-Fi, flash backup storage, and a USB Huawei E3531 3G modem. An mPower Ubiquiti (from here onwards referred to as mPower) device is used for refrigerator monitoring and control. As SSH is the primary means by which the mPower is controlled, the use of a GNU/Linux operating system makes its control by the Raspberry PI B+ trivial. SSH can also be used to connect to the FlexBox and download the data using SCP. An onboard storage microSD card on the Raspberry PI B+ makes data collection much simpler. If all other avenues fail to communicate the data to our server (an enumerator collecting data via Wi-Fi, or a 3G modem streaming data to our cloud server) the card can be mounted and read using a GNU/Linux based laptop. The modem is used to stream a subset of the data to our server, to control the FlexBox system, and to test the quality of the GSM network. The Ethernet port provides a fail-safe communications channel with the device. The software control system is developed using a combination of R, Python, and PosgreSQL. Bash scripts and CRON jobs are used to manage automated reboots, backups, and data archiving.

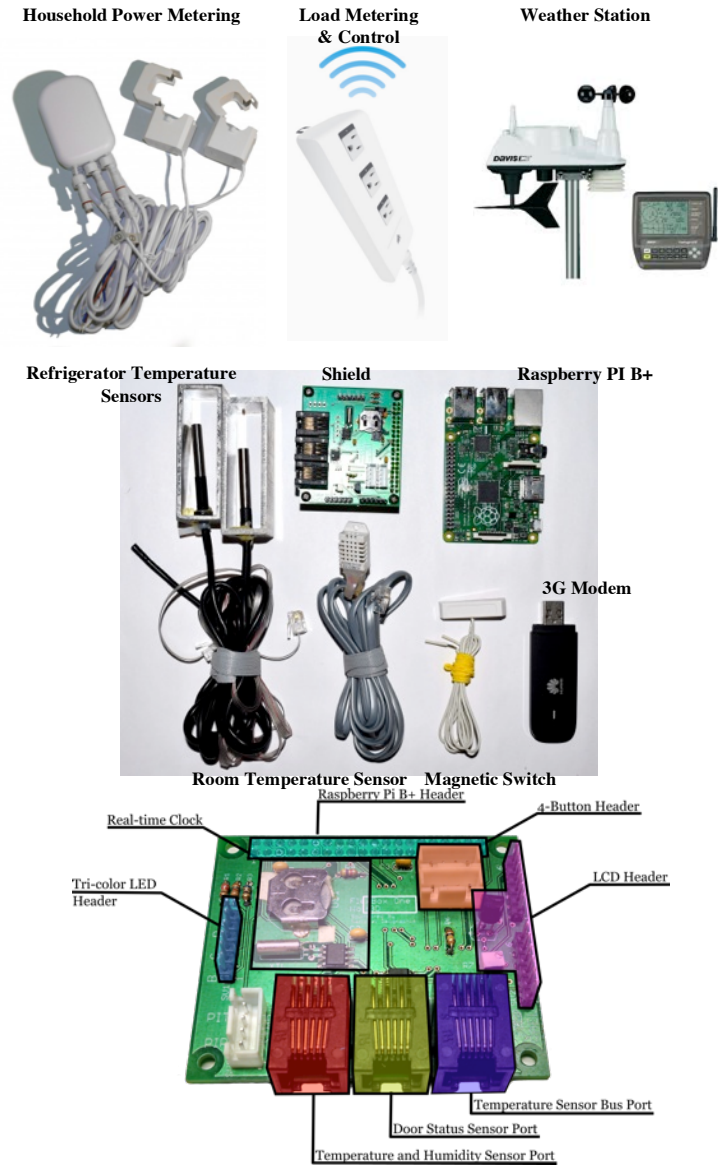

**Figure 2. FlexBox Wireless Sensor Gateway Components**

There are 7 sensors used in each FlexBox. Two DS18B20 waterproof temperature sensors (located inside the refrigerator) take measurements inside, a DHT22 temperature and humidity sensor monitors the room environment. A magnetically actuated reed switch monitors door openings. A mPower device controls the refrigerator and monitors power consumption. Finally, an Aeotec Power meter monitors house power consumption (located at the electric service panel). Several additions were made to the sensors and cables, including a small cage to surround the DS18B20s temperature sensors to minimize thermal contact conductance when inside the refrigerator as well as a thin telephone cable to extended the DS18B20s length (and allow for the refrigerator door to seal completely). FlexBox power and all sensors (connected via set of RJ11 modular jacks) come out through one of the sides of the box

Altogether there are 3 radios in the FlexBox design, WiFi, Z-Wave, and GSM. WiFi is used to communicate with the mPower load monitor and switcher, the Z-Wave dongle communicates with the Aeotec household clamp meter, and GSM is intended for remote data transmission and actuation control in a scenario where the FlexBoxes have a centralized controler. I2C is used to communicate with the temperature sensors and a proprietary protocol is used to communicate with the DHT22 room temperature and humidity sensor. Ethernet is also used for development purposes, which lets us download software packages and remotely login. Using higher-level development tools gives us the ability to iterate on changes to the software immediately reducing the iteration time to minutes.

### C. Software Implementation

The Raspberry PI B+ contains the Raspbian Wheezy operating system. All software is implemented in Python and all data is stored in a PostgreSQL database. The Raspberry PI B+ uses the host access point daemon (hostapd) to act as a WiFi access point in order to communicate with the mPower device. This access point also allows users to more easily connect to the FlexBox for diagnostics and data collection.

Values from switch sensor are directly accessible through the GPIO ports on the Raspberry PI B+. The temperature and room temperature and humidity sensors show up as character devices. An open-source repository called python-openzwave is leveraged in order to access voltage, current, power, power factor, and energy values from the meter. The python paramiko package is used to communicate with the mPower over WiFi through a secure shell (SSH) connection and collects refrigerator voltage, current, power, power factor, and energy values.

In order to maintain stable connections and handle communication errors, the data collection scripts incorporate several layers of connection and process resets. First, a separate process is created for the data collection script of each sensor. This allows for independent sensor reads and stores and prevents the failure of a single sensor from interfering with the collection of other sensor values. While this system is capable of collecting data every 1-3 seconds (depending upon the sensor), the limited storage capacity of the microSD card requires a more limited collection scheme. The software stores data in the PostgreSQL database under two conditions: 1) it detects a change in the output value that is greater than a specified threshold, and 2) one minute has passed. This second condition ensures that the sensor is still functional, otherwise it would be difficult to discern between a broken sensor and a static sensor output.

The sensors that communicate over wireless protocols (mPower, Aeotec Home Energy Meter) also have an additional layer of process handling to prevent excessive data loss caused by wireless connection issues. The Z-Wave network and connection to the mPower could be very sporadic. Each hour, or if any communication error is caught, the entire system process is restarted.

The mPower has an additional timeout for resetting the wireless network on the Raspberry Pi B+. If the Pi cannot connect to the mPower device, the wireless network is reset. After four retries, the Pi B+ stops attempting to connect and waits for the process to be killed in the subsequent hour. This limit was imposed to allow for users to access the Raspberry Pi's WiFi network even when the mPower is not functioning without having the script constantly resetting the connection.

The Python Flask microframework is used to set up a web server on the Raspberry Pi B+. A web page on this server allows users to easily see the last several data points that were entered into the database from each sensor. This allows for quick diagnostics by the enumerator when first entering a household. Other configuration properties include setting a static IP address for the mPower device, hard-coding the temperature sensor ids, and assigning unique hostnames to each FlexBox. These configurations add stability, reduce the possibility for error during system resets, and allow for easy identification and tracking when analyzing multiple households simultaneously.

Regarding communications, our approach seeks to evaluate two different network measurements: latency, and bandwidth. Latency represents the time interval in milliseconds between stimulation and response (how long it takes for data to get from one place to another), with bandwidth (bits per second) representing how much data can move across the network at a given time. Network latency is evaluated through pinging: every 30 seconds, 6 pings are sent which do a round trip to and from the server (FlexBox → cloud server → FlexBox). Bandwidth is measured every 2 hours by opening a transmission control protocol (TCP) connection with the cloud server and streaming 3 megabits of randomly generated numbers from the FlexBox (to prevent compression by the network which would inflate our perceived bandwidth). Every four hours one row of data (sensor and meter readings) is sent to the could server to update system parameters.

### D. Laboratory System Pilot

Our laboratory pilot tested the FlexBox to conditions that it might experience in the field. Laboratory testing was composed of baseline data collection, system probing, and system integration and calibration. A household refrigerator (Kenmore 253.68182800 top-refrigerator) from a common research space at the University of California Berkeley was instrumented with the FlexBox and baseline data was collected from the beginning of February 2014 to the end of April 2014. All data parameters were collected including door openings, fridge inside temperature, room temperature, refrigerator power consumption, and Aeotec measurements.

Communication challenges were the most restrictive during the testing period. With the FlexBox creating a dedicated WiFi network for monitoring and control, special permissions needed to be acquired to enable a new FlexBox network within the floor, and network stability proved to be a continuous problem. Initially, we used two Wifi Networks: a HT-TM02 TripMate Nano Router was used to connect to the University's WiFi network to stream second-by-second sensor data, and the USB Raspberry Pi's WiFi network was enabled to control the TCL (on/off) via the mPower. Data collected from all sources was stored locally as well as being streamed through the Univesity's WiFi network using BitTorrent Sync. Initially, a Huawei E220 3G modem was installed to enable bi-directional communications, but its spotty connectivity and power draw significantly affected system functionality. Later on, the same data tests were performed on a Huawei E3531 modem that is sold by telecommunication companies throughout Nicaragua (Claro). A fully functional FlexBox was probed with high humidity (90%) and high temperatures (40°C) without observing any operational impacts to weather shocks.

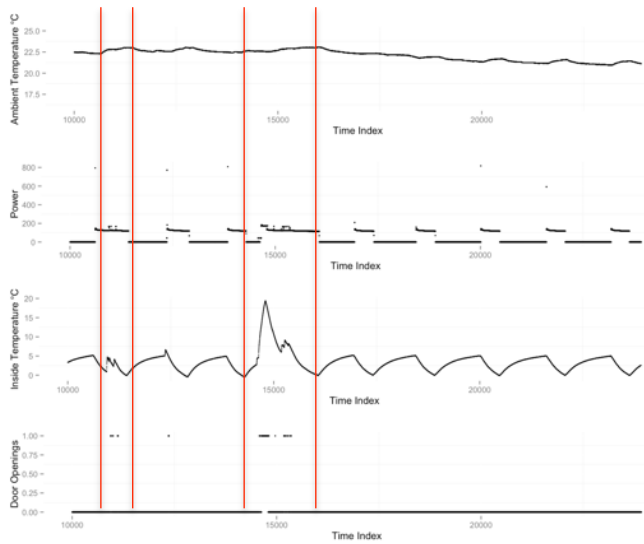

**Figure 3. Controlled System Pilot Data Stream:** Room temperature, power consumption, fridge inside temperature and door openings are all time stamped and temporally correlated. Small changes in room temperature can be observed, as well as the rapid increase in fridge inside temperature (and the longer time required to return to average cycling periods) after consequent door opening events (in red).

Even in the controlled environment of the University of California Berkeley's Sudartja Dai Hall, a building designed for energy efficiency and demand response experiments, our data began elucidating parameters that researchers could consider including when calculating the theoretical research potential of DR resources. For example, researchers have traditionally assumed that indoor air temperature is *constant* and *uniform* across households. Although small, we were able to observe an average daily variation of 3°C throughout the duration of our pilot, this variation was expected to be significantly larger in Nicaragua. The *constant* and *uniform* room temperature assumption is an important one as it also fixes other thermal parameters that allow for the calculation of theoretical DR resource potential, including: a constant duty cycle, fixed daily and seasonal mean power consumption ($P_m$), energy capacity ($E_{cap}$), and power capacity ($P_{cap}$) of an aggregation of TCLs. Under these assumptions one can estimate TCL dead band "width", temperature set points, thermal capacitance and coefficient of performance. Furthermore, previous research assumes that the effect of disturbances is small (door openings, for example) and can be ignored. In our controlled pilot we observed that while one short door opening might have relatively zero impact during a cooling cycle, continuous door openings (or one long door opening) can more than quadruple energy consumption (due to increased fridge inside temperature). We expect the FlexBox to accurately capture data that can allow DR researchers to make reasonable assumptions regarding the impact that weather, behavioral components, and energy efficiency, so as to build perhaps more realistic theoretical resource potential models.

V. FIELD DEPLOYMENT

In the summer of 2015 twenty micro-enterprises and ten households in different parts of Managua with similar social-demographic characteristics were selected at random from a sample of 300 micro-enterprises and households to receive a FlexBox. Five modems Huawei E3531 were installed to test network latency and bandwidth. This section presents an exploratory data analysis of the data collected to date, including TCL thermal parameter estimation and efficiencies, a brief communications network analysis, a cost breakdown, and a summary of field implementation challenges and opportunities that have been presented to date.

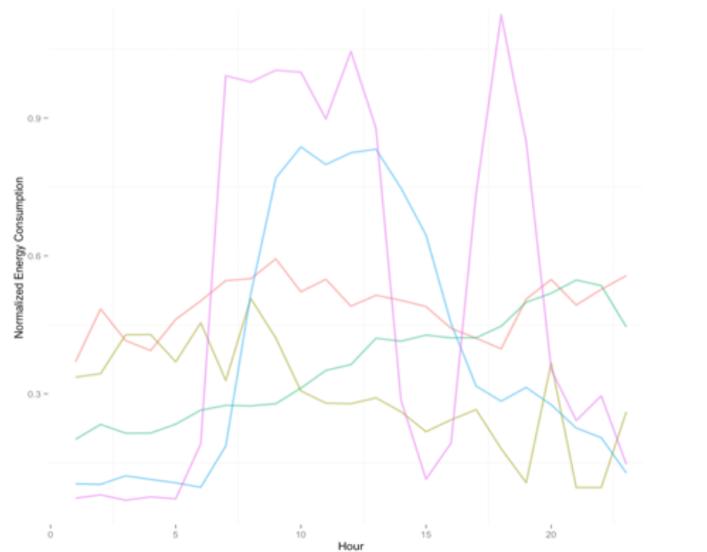

**Figure 4. Load Shapes:** Five different load shapes were identified when clustering load shapes by hourly mean or median. The pink load shape is the load shape that most resembles Nicaragua's characteristic daily demand load shape.

## A. Exploratory Data Analysis

Normalized micro-level energy consumption (mean and/or median) can be clustered into 5 different *daily load shapes* (hourly data). The five different clustered *daily load shapes* (k-shape) include (1) those that have their highest consumption in the middle of the day, (2) those with two peaks occurring in the middle of the day and in the evening, (3) those whose consumption increases consistently throughout the day, (4) those with only high consumption in the morning and at night, and (5) those that have scattered consumption throughout the day, but with the highest consumption being in the middle of the day. On average, households and micro-enterprises consume more energy on weekends versus weekends (mean: 16% greater energy consumption, median: 28% greater energy consumption using median).

Correlating time of day with hourly room temperatures (°C), fridge inside temperatures (°C), refrigerator energy consumption (Wh), and household energy consumption (Wh) allowed us to see that there is both a room temperature and time dependence (with varying correlation strength) across our cluster (Figure 5). We observe a very weak negative relationship between fridge inside refrigerator temperature and room temperature (Pearson r=-.06, N=16,000, p<.001), a moderate positive relationship between fridge energy consumption and room temperature (Spearman non-parametric r=-.43, N=10,000, p<.001), a moderate positive relationship between household energy consumption and room temperature (Spearman non-parametric r=-.45, N=16,000,

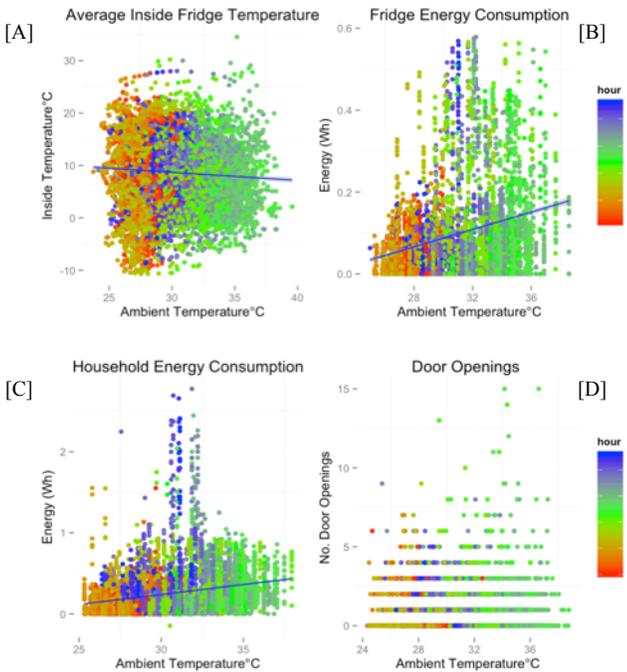

**Figure 5. Controlled System Pilot Data Stream:** Room ambient temperature plotted [A] fridge inside temperature, [B] fridge energy consumption, [C] household energy consumption, and [D] door openings. While the cluster only depicts weak to moderate correlations between room ambient temperature and other data streams, individual units experience stronger correlations between room ambient temperature and all other sensor data.

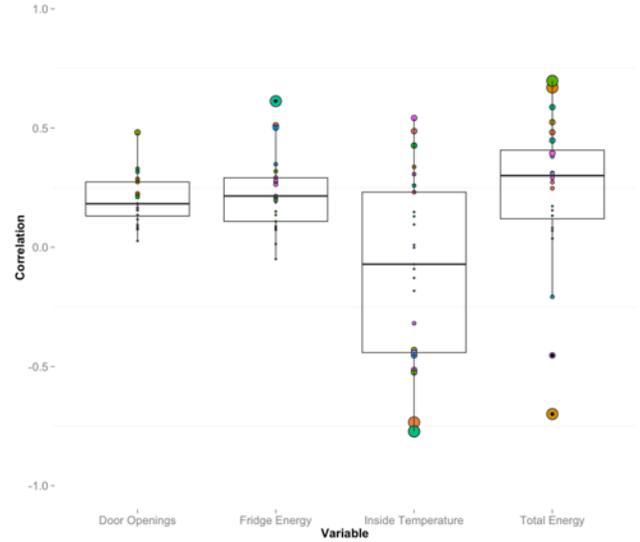

**Figure 6. Correlation between Room Temperature and FlexBox Sensor Data:** The figure depicts the strength of the correlation between room temperature and all other sensor readings (door openings, fridge energy consumption, fridge inside temperature, and total household energy consumption) for all units. *The size of a point represents the strength of the correlation and the color depicts a visual identifier for a specific unit (household or micro-enterprise)*. While the cluster data (all units) in Figure 5 depicts only weak to moderate correlations, individual units experience stronger correlations between room ambient temperature and all other sensor data.

p<.001), and a very weak relationship between room temperature and door openings (Spearman non-parametric r=-.16, N=16,000, p<.001).

While these relationships may be relatively weak across the cluster, we observe great within cluster variability when exploring these relationships (Figure 6). For example, Figure 6 depicts that 8 and 3 units experience moderate (0.4 ≤ r < 0.6; p<.001) and strong (0.6 ≤ r < 0.8; p<.001) correlations between room temperature and internal refrigerator temperature respectively, 6 and 3 units experience moderate (p<.001) and strong (p<.001) correlations between room temperature and house energy consumption respectively, and while the correlation between fridge energy consumption and room temperature is positive, this relationship is only found to be moderate and strong in a few households (2 and 1 units respectively; p<.001).

The spread in the strength of correlation between ambient room temperature and fridge inside temperature and fridge energy consumption suggests that there is a panoply of user behaviors that are driving the system (Figure 6). For example, some units might unplug their fridge when room ambient temperature is very high, whereas others might leave their appliance 'on', with the fridge using more energy to preserve (or reduce its internal temperature) during that time. Similarly a strong positive correlation between fridge inside temperature and room ambient temperature could suggest that users unplug their fridge during the hottest parts of the day, and a negative strong correlation could suggest that these are the times of the day when users actually 'plug' their refrigerator (and consequently, the time of the day during which the refrigerator uses most of its energy). The correlation between total household energy consumption and room ambient temperature suggests that while there are a few households that increase their consumption at higher temperatures, there are also others that modify their behavior so as to reduce their consumption (for example, turn several freezers and refrigerators off). There are many more insights from these data, including the opportunity to target energy efficiency thermal insulation for refrigerators in certain units, as well as the development of detailed energy reports.

### B. TCL Parameter EDA

There are several key parameters for determining the technical resource potential of thermostatically controlled loads and for building more accurate control algorithms for large-scale TCL aggregations. Room temperature, fridge inside temperature (of a room, or inside a refrigerator, for example), power consumption, and TCL characteristics (resistance, capacitance, and wall thickness, for example) are all used for the design of a smart controller. It has also been suggested that large-scale TCL aggregations of virtual energy storage can be represented through both their *energy* and *power capacity* [46]. To define the *energy capacity* (the maximum amount of energy that can be stored) and the *power capacity* (the full power range of an analogous storage device) several parameters are needed including: $h$ (the amount of time it takes a TCL to traverse its deadband in ON mode), dead-band width (°C), temperature set points (°C), thermal resistance (°C/kWh), thermal capacitance (kWh/°C), coefficient of performance (COP), and power consumption (kW). While TCL models in the literature allow room temperature to vary when modeling air conditioners and heat pumps, room temperature remains fixed when modeling energy and power capacity in refrigerators. Though these dynamics may vary across regions and study sites, a fixed room temperature also means that a refrigerator's duty cycle remains constant, and so do the power and energy capacities, as well as the mean annual energy consumption [46].

When comparing room temperature and humidity inside households and micro-enterprises against ambient weather station data, we found that houses and micro-enterprises directly experienced ambient temperatures, and often experienced hotter temperatures during the hottest part of the days due to the absence of reflective or insulating house materials infrastructure. During the early morning (0-6 am) all except two houses experience lower temperatures than the ambient temperature weather station, but this changes at 6 am when approximately half of the households experience higher temperatures than the weather station. While room temperature allows us to understand intra-hourly and intra day temporal variability, and temperature variability is well correlated across our weather station and all units, there was a wide spread of room temperature across all units (4°C). Poor thermal insulation could pose significant problems not only for household and city-wide energy efficiency programs, but could also significantly affect city dwellers health [47]. Our data suggests that not only does room temperature vary significantly during the day, but also that the warm temperature extremes are experienced significantly by loads and people in houses and micro-enterprises.

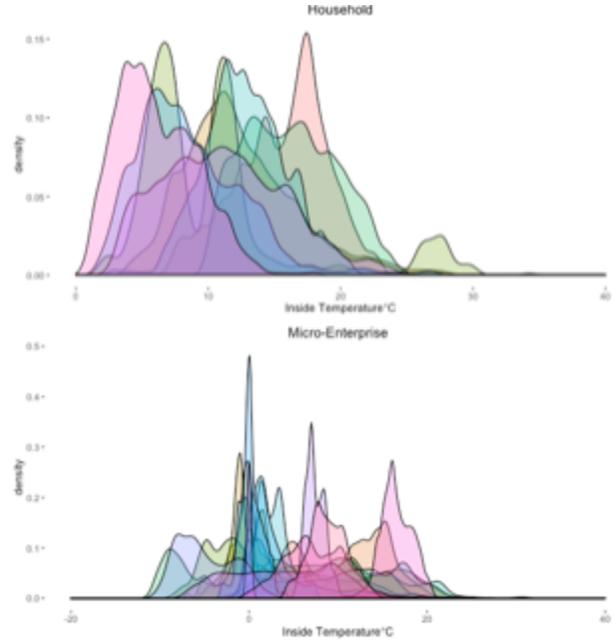

**Figure 7. Internal temperature of household and micro-enterprise TCLS:** The temperature dead-band of households is similar (top), while micro-enterprise freezers display a wider dead-band, ranging from -10°C to room temperature (bottom).

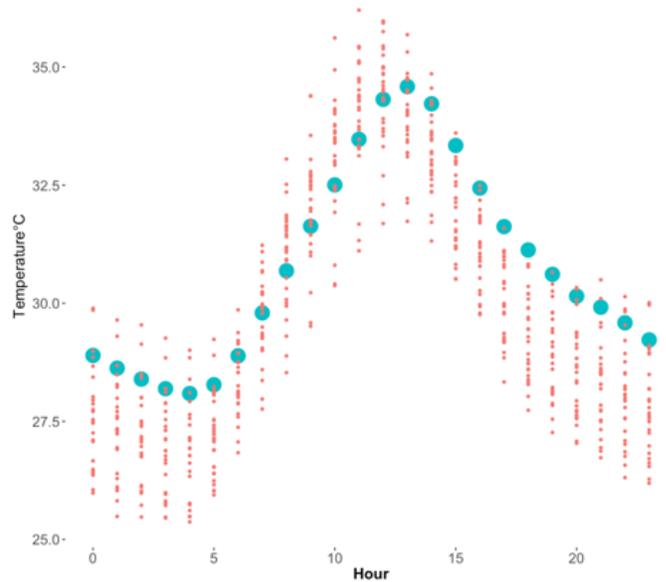

**Figure 8. Room temperature of household and micro-enterprises (red) vs. ambient weather station data (blue):** Houses and micro-enterprises directly experience the ambient weather station temperature on the streets of Managua. In the morning, late afternoon, evening and night, the weather station experiences higher temperatures than the households and micro-enterprises.

On average, TCL consumption is greatest during the middle day when it is the hottest and when households experience the majority of their door openings. Figure 8.a depicts normalized data (0-1) for all units to compare energy usage over time throughout the study period. Manufacturer information from refrigeration units in the field labeled the temperature set points of the different freezer and refrigeration units to range between -20°C and 5°C. Field data suggests, however, that the units usually oscillated between -10°C and could reach up to 35°C (Figure 7, Table 1). This deviation could be a result of appliance losses, and behavioral components which include the opening and closing of doors and the temporary unplugging of TCLs which most units engage in. Furthermore, we find that the duty cycle (the ratio of time it takes for a refrigerator to traverse its dead-band in an on state vs. total time in compressor on and off states) fluctuates during the day. Here, field data suggests that the freezers and refrigerators spend more time in the compressor-on stage during the middle of the day (when it's hottest and when there is more activity) than other parts of the day. Evidence from these field data diverge from previous TCL modeling assumptions that suggest that the duty cycle (and energy and power capacities) is fixed throughout the day.

We also compare the coefficient of performance, which was measured in an experimental setting at UC Berkeley, to an efficiency performance index, which was calculated from data. We find that while the experimental COP ranged between 0.01 and 0.03 and stayed fairly constant throughout the day (with minimal heat or behavioral disturbances), the efficiency of performance index (EPI) observed in the field ranged drastically between 0.0045 (minimum) and 18 (maximum). While it would seem like the EPI index is consistent across field units (Fig 8), we find that the performance efficiency of the refrigerator (the amount of work required to remove heat from a cold reservoir) changes during the day. More active and hotter times of the day observe lower EPI values than other days. The rated power of these appliances ranged from 0.1 to 2.2 kW according to the manufacturer label and size; this would result in a mean annual consumption range between 280 and 6000 kWh depending. Our field data suggests that the actual mean annual energy consumption was 1400 kWh for the entire cluster. Findings from our field data and experiment could be used to better inform the modeling of TCLs for ancillary services as theoretical models usually assume constant duty cycles, energy and power capacities and performance efficiencies.

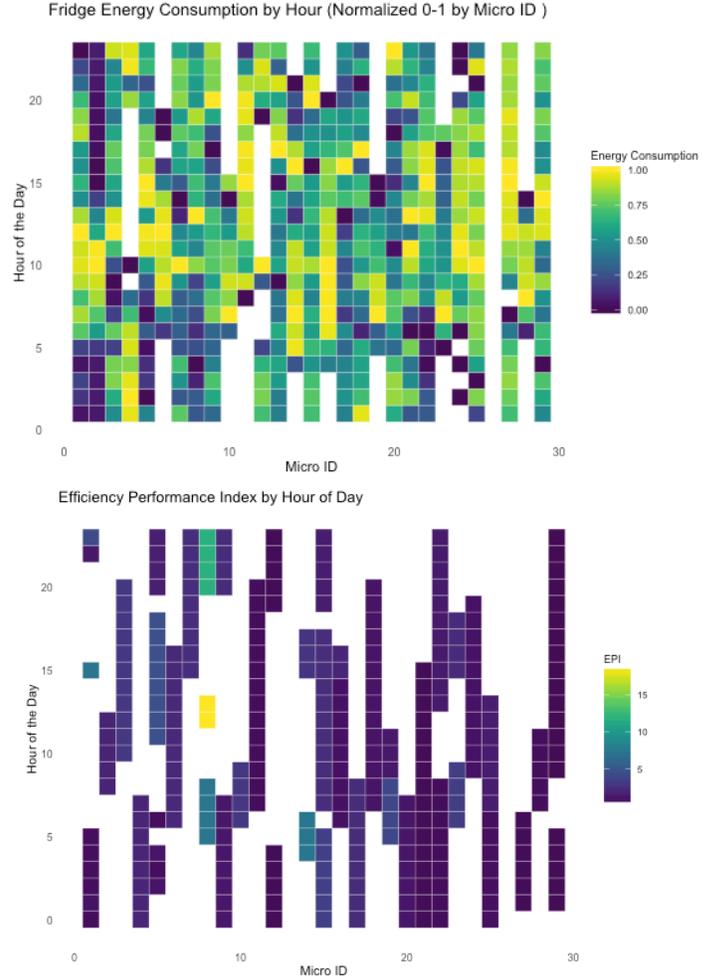

**Figure 9. Normalized TCL Energy Consumption by Unit [top] and TCL Efficiency Performance Index for all Units [bottom]:** [Top] We observe TCL energy consumption to be, on average, higher in the middle of the day than other parts of the day, and [Bottom] we find the efficiency performance index (the ratio between the work that is required to remove heat from a reservoir and the heat removed from a reservoir) also varies during the day, and is lowest in the middle of the day when it is hottest and when the TCL experiences most activity.

| Parameter | Symbol (Units) | Mean (SD: Min -- Max) |
|---|---|---|
| Ambient temperature | $\theta_a$ (°C) | 30 (3: 10 -- 41) |
| Dead-band width | $\delta$ (°C) | 9 (4: -10 -- 35) |
| Temperature set point[1] | $\theta_{set}$ (°C) | -20 -- 5 |
| Duty cycle | D (-) | 0.52 (0.31: 0.1 -- 0.9) |
| Coefficient of performance[2] | $\eta$ (-) | 0.01 - 0.03 |
| Efficiency performance index | $\eta.e$ (-) | 1.8 (2.4: .0045 - 18) |
| Power consumption[1] | P (kW) | 0.1 -- 2.2 |
| Mean Annual Energy Consumption per TCL[1] | MAEC (kWh) | 280 -- 6000 |
| Actual Mean Annual Energy Consumption per TCL | AMAEC (kWh) | 1400 |

[1] From product details found in the field and from local refrigerator and freezer providers.
[2] From controlled laboratory experiments. The literature suggests that the COP ranges between 1.5 and 2.5, we did not observe this in our controlled experiment. COP is a ratio of $Q_c$ (heat removed from a cold reservoir) over $W_{ref}$ (the work input required to remove heat from the cold reseroir). Experimentally, we calculated the COP for a freezer and refrigerator that were empty, but on the field we assumed freezers and refrigerators to be 3/4 full. That is, we used the heat capacity of air and water to calculate the efficiency performance index for our field data.
[3] The rest of the data was obtained from the field.

**Table 1. Field Data TCL Thermal Parameters**

## C. Communications Network EDA

As part of a test for the reliability and capacity of the communications network, we installed five 3G Huawei E3531 modems in both households and micro-enterprises. Monthly 1GB data plans were purchased for each modem and two tests were written and implemented to test network latency and bandwidth. Latency refers to the base overhead of establishing and responding to a connection request. In this context, it measures the amount it takes for the FlexBox to create a data package, send it to the server, the server receiving it and the server sending it back to the FlexBox. We measured latency through pinging: every 30 seconds, 6 pings were sent from the FlexBox to the server, and then returned back to the FlexBox. With regards to DR control purposes, latency is incredibly important as we want DR control signals to travel fast through the network. Modems can take anywhere between millisecond to tens of seconds to establish a connection and send one packet of data changing what service we can reliably provide in ancillary services or spot markets.

Bandwidth refers to the speed at which data flows through the network after a connection has been established and is usually taken into account when considering bulk data transmission. We measured bandwidth by opening a transmission control protocol (TCP) connection between the FlexBox and the server and transmitting 3 megabits of randomly generated numbers. For DR control purposes, control signals are generally very small and communication time is dominated by latency, so although we measured both, latency is considered to be a more determining factor of the ancillary services that could be provided by TCLs within a particular communications network. In our tests, and in the event that the network failed and latency and bandwidth data tests could not be sent, the tests were stored as failures in the FlexBox. Once the network was restored, data was sent to the server and analyzed to understand how many failures occurred based on how many sequence numbers were missing (as well as to calculate how much time had elapsed between successful attempts). Each latency and bandwidth test had 6 pings, and the maximum and average values described below refer to the maximum and average value within the 6 pings that occurred within each of our tests.

A non-parametric Kolmogorov-Smirnov test was used to compare the latency distributions across our five samples (for the average and maximum latency length) and found them to be all statistically significant different from each other (p≤0.001; with the null hypothesis being that the two distributions being compared are drawn from the same distribution). The mean of the average latency is 642 milliseconds (sd: 185 milliseconds) across all devices (Figure 10.A) and the mean of the maximum latency across all devices is 945 milliseconds (sd: 415 milliseconds) with the maximum latency value reaching 38,000 milliseconds. We also evaluated the average latency across all devices for every hour of the day (Figure 11.C) and found the network to be faster, on average, between 5 am and 12 pm (880 milliseconds) than other parts of the day. Distance between devices does not seem to be a determining factor of latency as devices that are relatively close together were found to be as different to each other compared to devices that were further away. The latency tests show great variability among each other and throughout the day although they are all connected to the same network (Claro 3G), are pinging the same server, are using the same technology (Huawei E3531) and run the same software.

We also analyzed network dropped packets and evaluated both the number of events (binary: 1 or 0) as well as the duration of the event (seconds: 1*seconds elapsed). Because our dropped packet events have both known average rates and are assumed to occur independently of the time since the last event, we assumed a Poisson distribution to express the probability of a dropped packet occurring within a fixed time interval. Similarly, we used an exponential distribution to describe the time between dropped packets (inter-arrival times of dropped packets in the Poisson process). Characterizing events as dropped packets within a Poisson process could allow us at a later stage to calculate the conditional probability of consecutive dropped packet events occurring within any given time period. Figure 10.A depicts the probability of an event (a dropped packet) happening within any given hour for each device. Each device has a different distribution, and therefore also a different lambda value. For all dropped packet events, the mean duration before reestablishing connection was 267 seconds (~ 5 minutes) with a 352 seconds standard deviation (min: 60 seconds, max: 206 minutes). These values are deceiving, however, because the distribution is skewed due to several extreme outliers shifting the mean to the right.

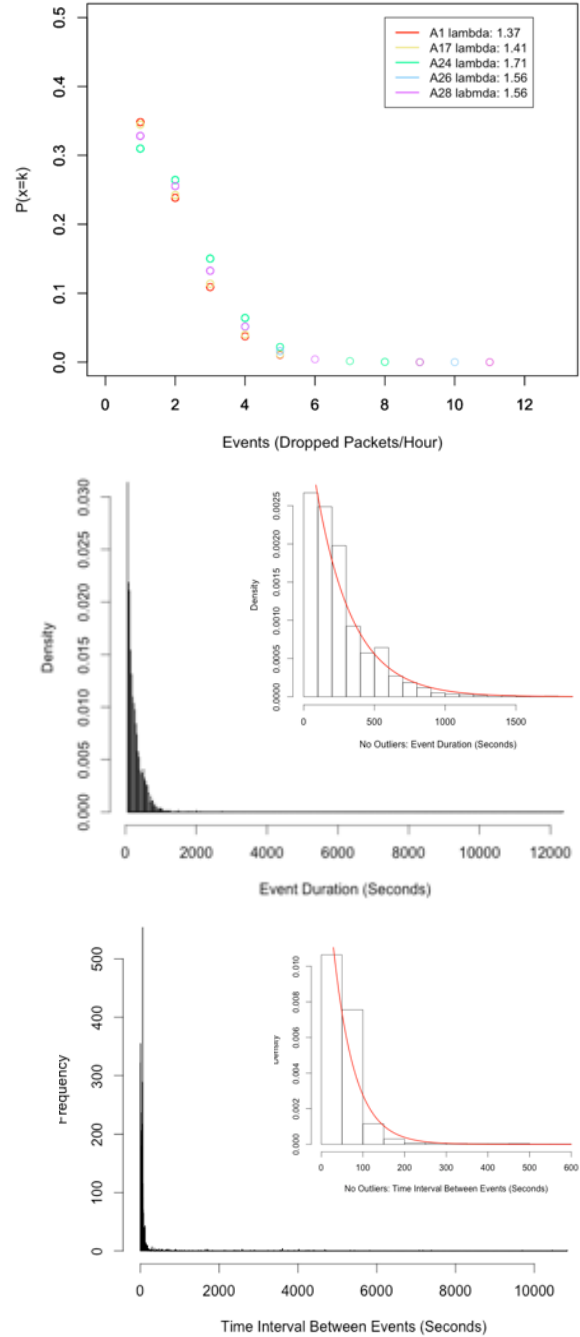

**Figure 10. Poisson and Exponential Distribution Characterizing:** Probability of number of dropped events per hour [A], distribution of event duration in seconds [B], and time interval between events [C]. Panels [B] and [C] depict the distributions without outliers and fitted with an exponential distribution (red line).

Removing these outliers depicts that the duration of events follows an exponential distribution with mean of 258 seconds (~4 minutes). Without outliers the median value is 180 seconds and the most frequent value is 60 seconds. Similarly, the time elapsed between events also follows an exponential distribution (Figure 10.C). For all dropped packet events the mean interval time between events was 106 seconds with a standard deviation of 505 seconds (min: 0.004 seconds, max: 180 minutes). These values change when removing extreme outliers. Without them, the mean time between events is 50 seconds (median is 46 seconds) with a standard deviation of 54 seconds. Future analysis will describe how these different event probabilities, event durations, and time intervals between events relate to the time when the grid most needs TLC resources to be available.

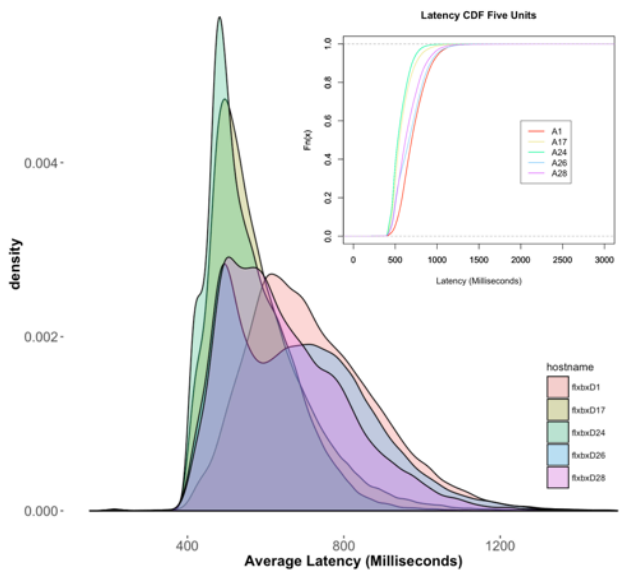

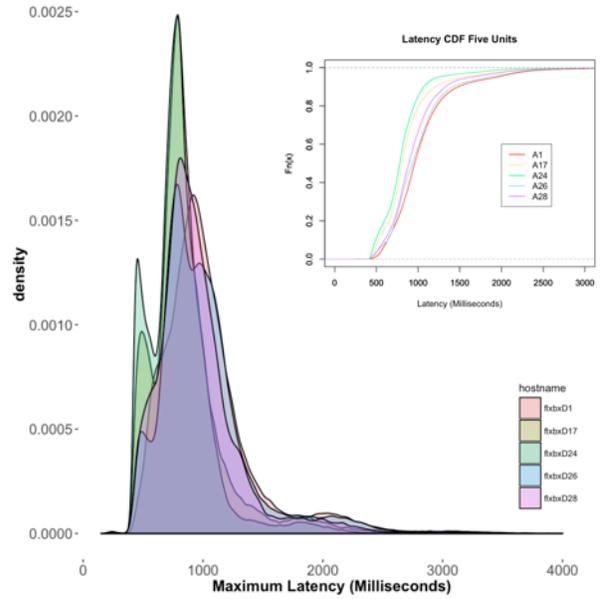

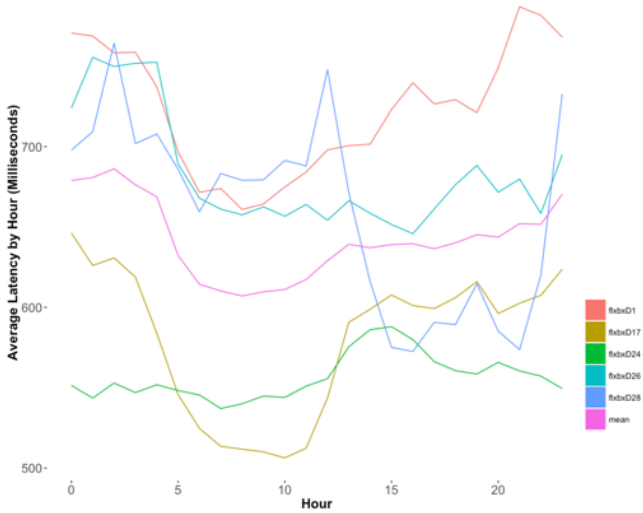

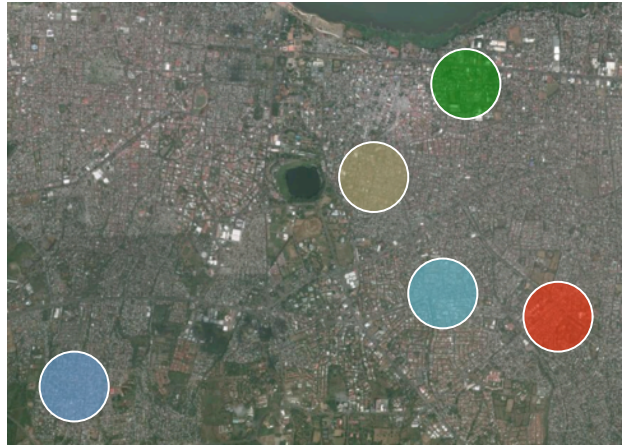

**Approximate Location of Devices across Managua**

**Figure 11. Latency and Bandwidth Distributions:** [A] Average latency, [B] Maximum latency, [C] Hourly average latency variability, and [D] approximate distribution of these devices across Managua.

## VI. CONCLUSION

The wireless sensor gateway implemented in Managua, and the data that is beginning to be collected and analyzed have the opportunity to inform how traditional models for DR could move forward. By complementing theoretical models with behavioral characteristics of users, resource uncertainty can be better characterized, and more attractive incentives for participation can be designed. There are several findings from our system implementation that can inform how theoretical models could incorporate data from wireless sensor gateways in the future: *(1)* a realization that modeling is a context specific endeavor, and thus, surveys and baseline data collection could be used for hypothesis development and assumption building before modeling begins *(assumptions that may hold in a particular context do not necessarily ought hot hold everywhere) (2)* while some recent work has begun to calculate the uncertainty resource potential for demand response, little attention has been placed on how user behavior increases the uncertainty on DR resource availability, *(3)* control algorithms are usually top-down with a load aggregator assuming user and load behavior and consumption patterns; we argue that a more holistic modeling approach could be the development of bottom-up – top-down models that incorporate behavior and appliance efficiencies in model building, *(4)* communication networks and enabling systems (such as our FlexBox) are usually discussed in the abstract, yet, the types of ancillary services that can be provided at the micro-level are conditional upon the capabilities of a specific system or technology, and *(5)* research on DR communication protocols are likely to affect not only what different services can be provided but also the design and cost-effectiveness of the enabling system itself. Future work will involve both the assumptions and model considerations explored in this paper, user and load characteristics, and a control algorithm for micro-level demand response.

As incomes and the ownership of loads increases together with the penetration of uncertain and variable renewable energy, grid management challenges will increase in tandem [1]-[7]. While the cost of AMI for large scale smart grid deployments may be cost prohibitive for many utilities in low, lower-middle income countries, 3G and 4G networks, as well as internet infrastructure (WiFi) will continue experiencing cost reductions and ubiquitous expansion in emerging economies [7][8]. Behind-the-meter IoT monitoring infrastructure could lead new ways of collecting and analyzing data for grid management that are mutually beneficial for grid operators and users. IoT grid sensing takes different design considerations into account including user acceptance, affordability, and assurance that the acquired data has benefits both for the user and as well as the load aggregator, or grid operator. AMI or smart meters, on the other hand, were not designed with consumer acceptance in mind. Designing IoT infrastructure for RCEs has the added benefits of both reverse innovation learning and leap frogging. The former allows high-income countries (e.g., California, Denmark) to learn from cost-effective innovations and deployments in RCEs to potentially reduce costs in their own countries, while the latter allows RCEs to skip technology iterations that could be sub-optimal.

Future iterations of this work will involve the reduction in size of the FlexBox, design of a system that measures temperature less intrusively, and a more inconspicuous way to measure load power consumption. In addition, future work will investigate what the minimum level of grid sensing is to recover full state information from a micro-enterprise or a household. Furthermore, although this work has described the system that can enable micro-level DR in resource constrained environments, the analysis necessitates the coupling of all this information with grid data in order to understand the resource that is available for DR resources taking into account both thermal parameters and the inherent challenges that exist in the communications network.

## Acknowledgements

CONACYT-UCMEXUS, the National Geographic Energy Challenge, the Berkeley Energy and Climate Institute, and UC Berkeley's Development Impact Lab partly funded this research. Nicaragua's National Engineering University, and Nicaragua's Ministry of Energy and Mines were institutions that actively supported the pilot implementation. We would like to thank Duncan Callaway, Emerald Ferreira-Yang, Jae Bin Ju, Odaly Maria Molina Altamirano, Jorli Jarqui, and Maria Virginia Moncada for valuable conversations and insights during the design and implementation of this research project.